# Detecting fatigue in aluminum alloys based on internal friction measurement using an electromechanical impedance method

Jihua Tang[#], Mingyu Xie[#], Faxin Li[*a)]

LTCS, Department of Mechanics and Engineering Science, College of Engineering, Peking University, Beijing, 100871, China

**Abstract**

Detecting mechanical fatigue of metallic components is always a challenge in industries. In this work, we proposed to monitor the low-cycle fatigue of a 6061 aluminum alloy based on internal friction (IF) measurement, which is realized by a quantitative electromechanical impedance (Q-EMI) method using a small piezoelectric wafer bonded on the specimen. Large strain amplitude ($3.3\times10^{-3}$) was employed thus the fatigue life can always be below $10^5$ cycles. It was found that except for the initial testing stage, the IF always increases steadily with the increasing fatigue cycles. Before the fatigue failure, the IF can reach 2.5 to 3.4 times of the initial value, which is thought to be caused by the micro-cracks forming and growing. In comparison, the resonance frequency of the specimen just drops less than 2% compared with the initial value. Finally, a general fatigue criterion based on IF measurement is suggested for all the metallic materials.

---
[#] Equivalent contributions
[a)] Author to whom all correspondence should be addressed, Email: lifaxin@pku.edu.cn

1. **Introduction**

Metallic structures are mostly widely used in modern industries and statistics has shown that over 80% of their failure is caused by mechanical fatigue. The fatigue problems of metals had been investigated over one century. However, so far the fatigue status of engineering structures still cannot be identified and their remaining life cannot be predicted yet. In practice, to ensure the safety of equipment, the operation period is usually fixed and the equipment is forbidden to work out of this period even if its condition is still quite good. Furthermore, routine nondestructive testing is usually required to detect the possible damage during the operation period. However, currently none of the existing NDT methods can effectively detect the micro-cracks generated at the early stage of mechanical fatigue.

In the past decades, intensive efforts had been made to study on the NDT methods for fatigue damages, among which the nonlinear ultrasonic testing (NUT) method may be the most well-known[1-3]. NUT had been shown to be sensitive for low cycle fatigue of metals, but less sensitive for high cycle fatigue. Actually, the big problem of the NUT is that it is strongly dependent on the testing equipment, driving power, couplant, etc, which prevents it to be a standard method. Ultrasonic attenuation had also been used to evaluate the fatigue in metals[4-6]. The early results by Joshi and Green[4] had shown that the ultrasonic attenuation coefficient α can increase 10% to 20% at about 90% of the fatigue life, which should be caused by the wave reflection at the crack faces. However, such a small increase in α makes little sense in practical applications. Ogi et al studied the ultrasonic attenuation coefficient α of surface waves in different metals during torsional-bending fatigue[5, 6]. They found that the in-situ measured α can rise to three to four times of the initial value before failure. However, the interrupted measurement shows that α does not change much at all. These results indicate that the ultrasonic attenuation coefficient α is insensitive to micro-cracks and thus not suitable for fatigue testing.

Actually since 1950s, physicists had studied the relationship between internal frictions

(IF) and fatigue in common metals including copper, aluminum, etc[7-13]. The pioneer physicist, W.P. Mason, developed an ultrasonic fatigue testing machine which can measure the amplitude dependent internal frictions during fatigue testing[7]. He found that before fatigue failure, the IF will increase rapidly and the Young's modulus drops quickly. Fred and James[8, 9] studied the internal friction variations in polycrystalline copper and aluminum subjected to irradiation and fatigue. In their works, wire specimens were used and the IF is measured by using the inverted torsional pendulum. They found that the IF significantly drops after irradiation, and before the fatigue failure, IF also increases rapidly. Kenawy et al[12] observed the IF of aluminum alloy wires will increase significantly immediately after fatigue deformation, but drops almost to the initial values after 10 mins. Bajons et al studied the amplitude independent and amplitude dependent internal frictions (AIIF and ADIF) in polycrystalline copper after ultrasonic fatigue using the torsional pendulum method and resonant bar method[10, 11]. They found that both types of IFs dropped after fatigue deformation. But in their works, the specimens did not fail after fatigue testing.

From above, it can be seen that the internal friction (IF) of metals is very sensitive to fatigue deformation. However, previously employed IF measurement methods, i.e., the torsional pendulum method, the resonant bar method, DMA[14], etc., can only be applied to specific shaped specimens and the measurement repeatability is not good, thus they are not applicable for practical structures.

In this work, we employed our recently proposed, electromechanical impedance based IF measurement method (called Q-EMI or M-PUCOT)[15, 16] to monitor the IF variations in a 6061 aluminum alloy during low-cycle fatigue. The Q-EMI method can be applied to any shaped specimen, and the IF measurement is highly repeatable and very quick. Results show that during fatigue loading, after the initial drop, the IF increases steadily with the loading cycles. Before fatigue failure, the IF can reach about 2.5 to 3.4 times of the initial values. In comparison, the resonance frequency of the specimen drops only about 2%. A general fatigue criterion based on IF measurement is thus proposed, which is applicable for all the metallic materials. This work is expected to pave the way to fatigue prediction of engineering structures or components.

## 2. Experimental

2.1 The IF measurement method

In this work, our previously proposed IF measurement method[15], is used to monitor IF variations during fatigue. This method is called quantitative electromechanical impedance (Q-EMI) method or modified piezoelectric ultrasonic composite oscillator technique (M-PUCOT). We prefer to the first name here since this method is not limited to ultrasonic range, it can work well even in the range of ~kHz. Meanwhile, the quick electromechanical impedance measurement is a big advantage of this method. The measurement principle of the Q-EMI method is illustrated in Fig.1.

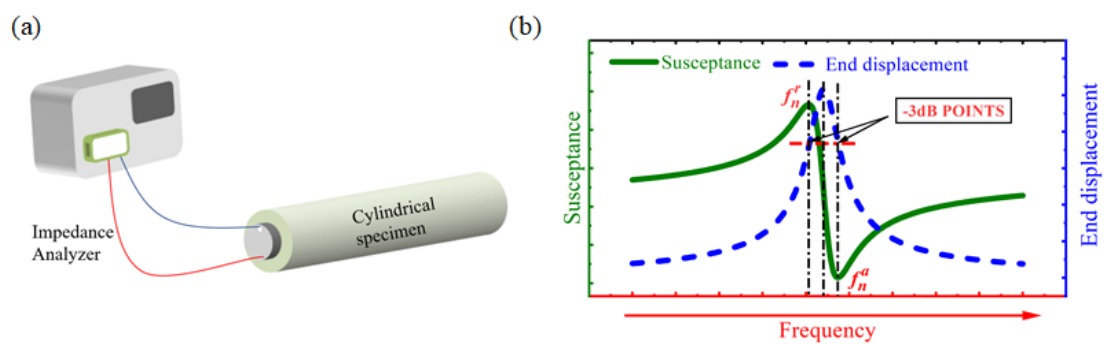

Fig.1 The internal friction measurement principle of the Q-EMI method. (a) Cylindrical specimen with a bonded piezoelectric transducer; (b) typical susceptance curve (green, solid) and the vibration displacement amplitude curve (blue, dashed).

As shown in Fig.1(a), a cylindrical specimen is used to show the measurement principle. A thickness poled PZT wafer is bonded on to the end of the specimen, forming a two-component vibration system. The length of the specimen is much larger than the thickness of the PZT (with the length to thickness ratio typically larger than 100), thus the resonance frequency of the two-component system (~ tens of kHz) is very close to that of the specimen, and much lower than the resonance frequency of the PZT wafer (~1MHz). A typical susceptance curve of the two-component system is shown in Fig.1(b) (green and solid curve), from which the resonance frequency and the anti-resonance frequency can be clearly identified. For comparison, the vibration amplitude curve of the two-component system is also plotted in Fig.1(b) (blue and dashed curve). Based on Ref.15, the resonance frequency $f_n^r$ and the anti-resonance frequency $f_n^a$ correspond to the two -3dB points of the displacement amplitude curve (blue and dashed curve)[15]. The mechanical resonance frequency of the two-component system is exactly the mean value of $f_n^r$ and $f_n^a$. The internal friction and mechanical resonance frequency of the two-component system can then be calculated by the following formula [15]:

$$Q^{-1} = 2\frac{f_n^a - f_n^r}{f_n^a + f_n^r} \quad (1)$$

$$f_{resonance} = \frac{f_n^a + f_n^r}{2} \quad (2)$$

As the volume of the PZT wafer is much smaller than that of the specimen (typically below 1%) and the resonance frequency of the two-component system is far below that of the PZT wafer, the IF contribution by the PZT wafer can be neglected here and Eq.(1) is employed to calculate the IF of the specimen. It should be noted that Eq.(1) is also applicable to other shaped specimen as long as the volume of the specimen is much larger than that of the PZT wafer.

2.2 Low-cycle fatigue testing

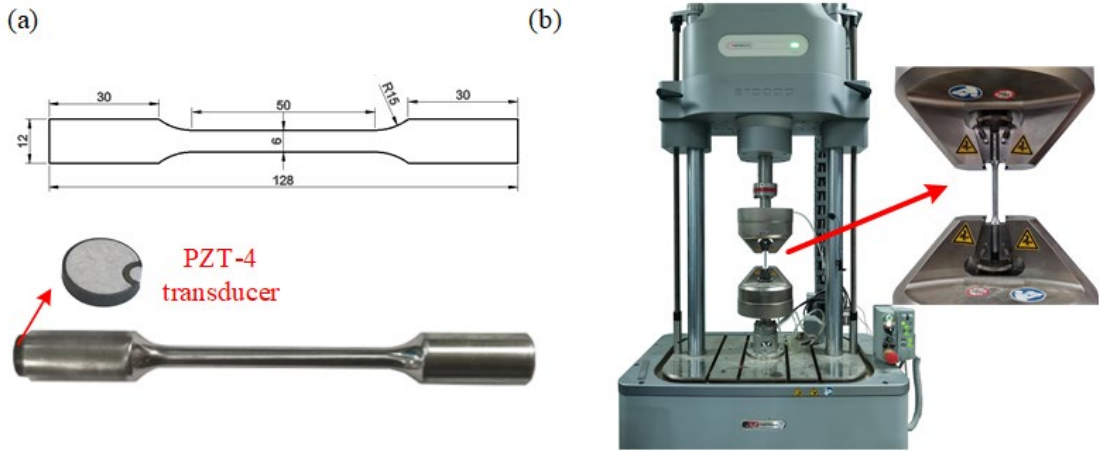

Fig.2 (a) 6061 aluminum specimen and (b) experimental setup for the tension-compression fatigue testing

The standard dog-bone shaped 6061 aluminum alloy specimen was used for the low-cycle tension-compression fatigue testing. The dimensions of the specimen are shown in Fig.2(a). Before the fatigue testing, tensile testing was conducted on the specimen to determine the yielding stress and Young's modulus. Then, all the specimen for fatigue testing were boned with a PZT wafer at one end for the IF measurement, seen in Fig.2(a). Finally, low cycle tension-compression fatigue testing was conducted on the 6061 aluminum specimens using an electromechanical fatigue testing machine (INSTRON E10000). The photography of the whole testing setup is shown in Fig.2(b).

The strain amplitude used in the fatigue testing is $3.3\times10^{-3}$ with the stress amplitude of about 230MPa, and the fatigue life under this strain level can always be below $10^5$ cycles. The fatigue testing were frequently interrupted for the IF measurement, typically with the interval of 3000 cycles. The susceptance curve measurement were conducted immediately when the fatigue testing was interrupted, and the stabilized admittance curve after 10 mins were used for the IF calculation.

## 3. Results and Discussions

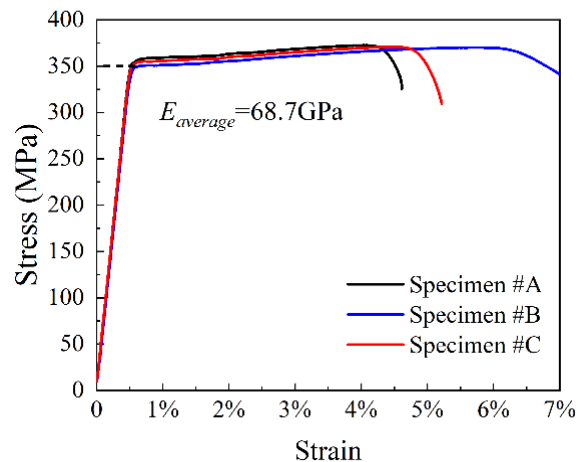

Fig.3 The tensile stress-strain curve of 6061 aluminum specimen

Fig.3 shows the tensile stress-strain curves of three typical 6061 aluminum specimens. It can be seen that the yielding stress is about 350MPa and the Young's modulus is about 68.7GPa. Then the loading stress amplitude of 230MPa is 65 percent of the yielding stress and the fatigue life can be controlled in the low-cycle fatigue range.

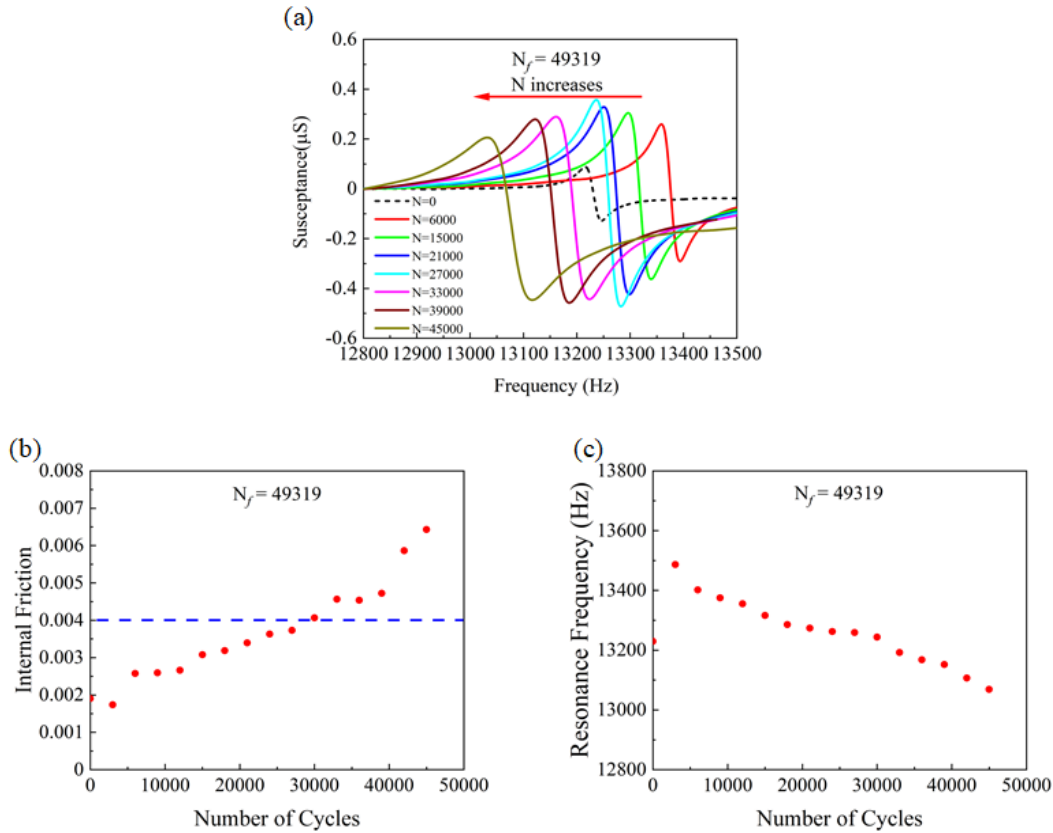

Fig.4 Measured parameters of the 6061 aluminum specimen #1 after different fatigue loading cycles. (a) the susceptance curve; (b) internal friction; (c) mechanical resonance frequency.

A series of 6061 aluminum specimens were tested and the results of three typical specimens were presented for discussions. Fig.4 shows the measured results of the Specimen #1 with the fatigue life of 49,319 cycles, where the susceptance curves, the internal friction and the mechanical resonance frequency after different loading cycles were presented, respectively. It can be seen that except for the case before loading, the internal friction of the specimen increases steadily with the loading cycles. The IF after 3,000 cycles is 1.7E-3, while the IF just before failure (after 45,000 cycles) reaches even 6.5E-3. Note that the initial IF of Specimen #1 before loading is about 1.9E-3, and after the first 3000 cycles of loading, the IF slightly drops to 1.7E-3. This tendency is consistent with the early measurement results on pure aluminum by Fred and James[9] and the slight drop of IF is attributed to the dislocation pinning. The subsequent steady increase in IF should be caused by micro-cracks forming and growing. From Fig.4, it can be seen that the IF of 6061 Al is very sensitive to fatigue loading.

The variation of the mechanical resonance frequency of Specimen #1 is just opposite to that of the IF, as seen in Fig.4(c). During the first 3000 cycles loading, the resonance frequency firstly increases from the initial value of 13,230 Hz to 13,486 Hz. After that, it drops steadily with the loading cycles, arriving at 13,068 Hz after 45,000 cycles of loading. It should be noted that the variations in the mechanical resonance frequency during fatigue loading is very small, about 3% in Specimen #1. Therefore, the resonance frequency is not a good measure for fatigue detection.

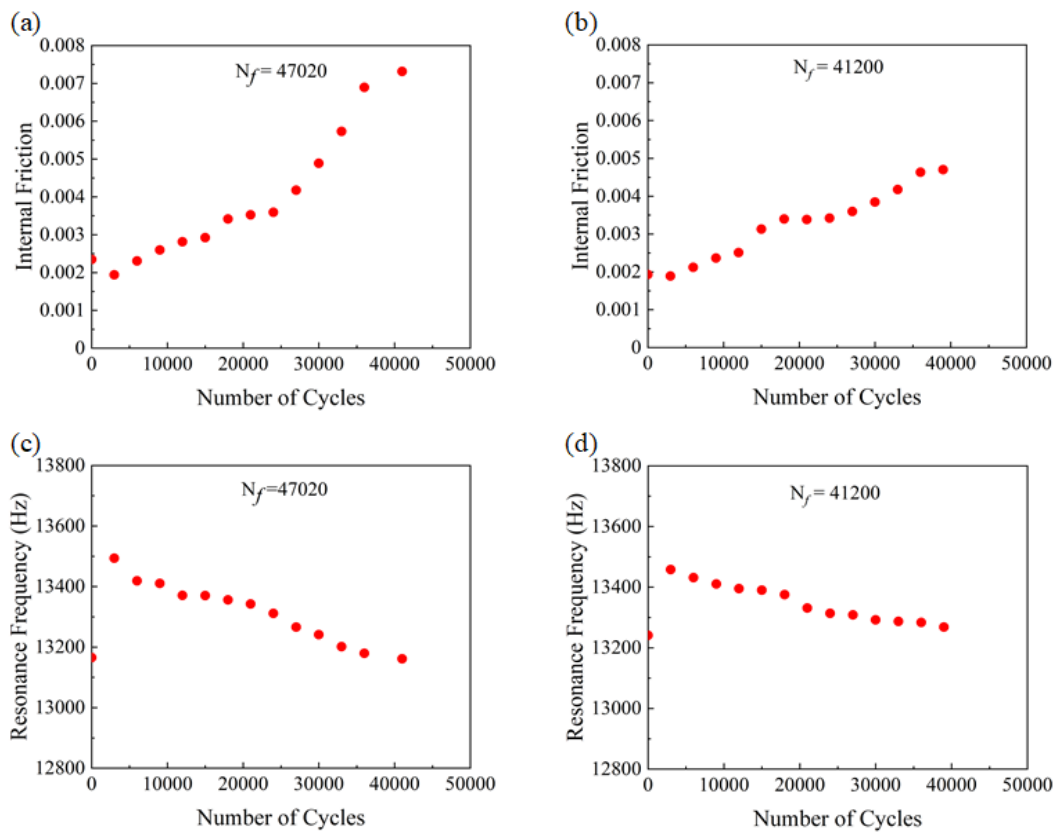

Fig.5 Measured internal frictions (up) and mechanical resonance frequencies (down) of aluminum Specimen #2 (left) and Specimen #3 (right) after different fatigue

loading cycles.

Fig.5 shows the measured internal frictions and mechanical resonance frequencies of another two specimens (#2 and #3) after different cycles of fatigue loading. The fatigue life of Specimen #2 and Specimen #3 is 47,020 and 41,200, respectively. The measured results are similar with that of Specimen #1, i.e., the IF firstly drops slightly then increases steadily with the loading cycles, and the resonance frequency is insensitive to the loading cycles.

Table I Measured IF of three 6061 Al specimens during fatigue loading

| Specimen No. | Fatigue life | Initial IF | IF after 3k cycles | IF before failure | IF before failure / Initial IF |
|---|---|---|---|---|---|
| #1 | 49,319 | 1.9 E-3 | 1.7 E-3 | 6.4 E-3 | 3.37 |
| #2 | 47,020 | 2.3 E-3 | 1.9 E-3 | 7.3 E-3 | 3.17 |
| #3 | 41,200 | 1.9 E-3 | 1.9 E-3 | 4.7 E-3 | 2.47 |

The measured IF of the three 6061 Al specimens during fatigue loading were listed in Table I. The ratio of the IF before failure to the initial IF is also presented, which varies from 2.5 to 3.4 for the three specimens. Note that the variations in the initial IF of the three specimens (1.9-2.3E-3) should be mainly caused by the different process states. The different PZT transducers and bonding conditions can have some effects on the measured IF, but this effect can be controlled within 1.0E-4 for the dog-bone shaped specimen in this work and can even be as small as 2.0E-5 for a cylindrical specimen[15].

*Suggested fatigue criterion based on IF measurement*

From above, it can be seen that for the 6061 Al during low-cycle fatigue, the IF is very sensitive to fatigue loading. Except for the very early stage of loading, the IF increases steadily with the loading cycles and can reach 2.5 to 3.4 times of the initial value. Therefore, the IF can act as a good indicator of fatigue damage and a fatigue criterion based on IF measurement can be established. This sounds reasonable since all the fatigue failure is accompanied by micro-cracks forming and growing, which will significantly enlarge the internal frictions.

The fatigue criterion can be based on absolute IF value or relative IF. For example, if we take "IF=4.0E-3" as the fatigue criterion, then based on Fig.4 and Fig.5, the warning lives for the three specimens are 30000, 27000 and 32000 cycles, which corresponds to about 61%, 57% and 78% of the whole fatigue lives. Alternatively, if we take "the relative IF (Measured IF / Initial IF) =2.2" as the fatigue criterion, then the warning lives of the three specimens turn to be 31000, 31500 and 33000 cycles, which corresponds to about 63%, 67% and 80% of the whole fatigue lives. The fatigue criterion based on the relative IF seems better since the IF variations among different specimens can be avoided. Obviously, for those central components, a lower relative IF, such as 2.0 or lower, is suggested to reduce the risk of fatigue failure. In addition, although this fatigue criterion is built based on low-cycle fatigue testing of aluminum, it should be applicable for all the structural metals under both low-cycle and high-cycle fatigue, since in these cases micro-cracks forming and growing always take a long time before the fatigue failure.

## 4. Conclusions

In summary, we monitored the internal frictions (IF) of 6061 Al during low-cycle tension-compression fatigue loading based on an electromechanical impedance method using a PZT wafer bonded on one end of the specimen. The measured IF firstly drops slightly then increases steadily with the loading cycles. Before the fatigue failure, the IF can reach 2.5 to 3.4 times of the initial value. In comparison, the resonance frequency drops less than 2% before failure. A fatigue criterion based on IF measurement is proposed and initial examination indicates that the relative IF is better than the absolute IF here. This work is expected to pave the way to fatigue prediction of metallic components and structures.

## Acknowledgements

This work is supported by the National Natural Science Foundation of China under Grant No.

12172007.

**DATA AVAILABILITY STATEMENT**

The data that support the findings of this study are available from the corresponding author upon reasonable request.